# Digital Accessibility Literacy

A Conceptual Framework for Training on Digital Accessibility

Björn Fisseler
Faculty of Psychology
FernUniversität in Hagen
Hagen, Germany
bjoern.fisseler@fernuni-hagen.de

## ABSTRACT

Developing digital accessibility expertise is critical to breaking down barriers and ensuring digital inclusion. However, a discourse on a pedagogical culture for teaching digital literacy is still lacking. This article, therefore, takes up the current discourse on the description of literacy and uses it to develop the concept of digital accessibility literacy as a fundamental element for promoting a pedagogical culture of digital accessibility. Digital accessibility literacy encompasses both the creation (encoding) and interpretation (decoding) of accessible digital content and technologies. By integrating awareness, technical standards, inclusive design practices, and continuous feedback into curricula, future professionals will be empowered to create digital environments that are accessible to all. This comprehensive approach improves technical skills and instills ethical and social responsibility. As a first draft of a digital accessibility literacy concept, the proposal will be used as a basis for discussion and further development.

## CCS CONCEPTS

• Computing education > Model curricula  • Human-centered computing > Accessibility theory, concepts and paradigms

## 1 Introduction

As digital technology becomes more integrated into everyday life and digital learning is pervasive in education and lifelong learning, the importance of digital accessibility and inclusion has grown significantly. This shift highlights the need for professionals skilled in creating digital content that is accessible to everyone, including people with disabilities.

Digital accessibility in different disciplines ensures that all users can fully engage with and use products, services, information, and environments regardless of their abilities.

To meet this demand, there is an urgent need to focus on training, education, and skills development for professionals from various disciplines in digital accessibility. This includes not only teaching the technical skills needed to create accessible content but also fostering an understanding of the principles of inclusion and accessibility.

Existing educational approaches to teaching digital accessibility must be critically evaluated and expanded to keep pace with technological advances and societal needs. This paper introduces a new concept called "Digital Accessibility Literacy" as a comprehensive framework for equipping individuals with the necessary knowledge and skills. By exploring current theories and practices, the idea aims to improve the effectiveness of education and teaching in this crucial area, ensuring that future professionals are well-prepared to contribute to a more inclusive and diverse digital world.

## 2 Teaching Accessibility in Various Disciplines

### 2.1 Teaching Accessibility in Computer Science

Teaching digital accessibility skills in computer science has been discussed in the US for several years. Several studies [5, 6, 13] showed that although accessibility is considered necessary by the teachers surveyed, only 15% of all teachers actually teach digital accessibility [5]. In a study by Shinohara et al. [13], 52% of respondents stated that the topic was not part of the curriculum. To establish digital accessibility in computer science curricula, Shinohara et al. [13] suggest integrating accessibility as a cross-cutting theme in the different sub-disciplines of computer science.

For Lewthwaite and Sloan [7], learning digital accessibility is more than teaching technical and technological skills. They consider digital accessibility to be a socio-technical challenge that is primarily about the problem of teaching empathy [7]. A "good practice problem" can currently be observed. Many stakeholders communicate the issue of digital accessibility in the same way in terms of content and methodology because everyone else is doing the same. For example, when universities inform their faculty about accessible Word documents, this must be the right way - because everyone else is doing it. In response, Lewthwaite and Sloan [7] want to develop a pedagogical culture of teaching digital accessibility and encourage greater sharing and interdisciplinarity.

To better integrate accessibility into the computer science curriculum, [2] present an Accessibility Knowledge Area that outlines important topics as so-called Knowledge Units. The topics of the units seem to follow the development processes, starting with "Disability Awareness", moving from the basics of "Accessibility Design" to "Accessibility Implementation" and "Accessibility Evaluation", and ending with "Accessibility Profession and Continuous Learning". For each proposed unit, the authors propose a list of topics along with learning objectives. The overall goal is to move the topic of accessibility beyond HCI and to integrate it into the entire computer science curriculum.

## 2.2  Teaching Accessibility in Pedagogy

For digital learning and teaching to be accessible, prospective teachers and other pedagogical experts also need appropriate skills. Thus, various models emerged in recent years to describe what teachers need to learn and know about the digitalization of teaching and learning and digital accessibility. One such model is the Technological Pedagogical Content Knowledge Model (TPACK) [4]. It describes competencies in different, partly overlapping areas of knowledge. It is based on the assumption that teachers must have content and pedagogical and technical skills to use digital technologies in teaching and learning processes successfully. At its core, the TPACK describes technological, content-related, and pedagogical competencies and combines these in various facets to form a competence model in which the three competence areas combine to create the TPACK. Though the original TPACK does not address digital accessibility, there are also ideas for expanding the model to include aspects of digital accessibility and inclusion [8]. Other, more broadly based competence descriptions are the European Framework for the Digital Competence of Teachers (DigCompEdu) [11] and the UNESCO ICT Competency Framework for Teachers [14]. The DigCompEdu competence framework divides teachers' digital competencies into six competence areas with 22 different competencies. In the English-language version, the competence area "Empowering Learners" includes the competence "Accessibility and Inclusion". Teachers should have the skills to consider accessibility issues when selecting, adapting, and creating digital teaching and learning resources and to improve accessibility [11].

## 3 Concepts of Literacy in Teaching and Education

The term "literacy" originates in the English language and refers narrowly to individual reading and writing skills. In 1997, Paul Gilster introduced the term of the same name in his book "Digital Literacy" [3] to describe the ability to understand and use information in various forms and from multiple sources when presented via the computer. However, there is more to the concept of digital literacy than simply reading information on the internet. "Reading" also means understanding a problem and developing a set of questions to address and solve the problem. This requires developing search skills to find relevant information online, critical thinking to make informed judgments on the information one finds online, and targeting your reading to solve your initial problem. As new technology is never neutral but favors certain groups while hurting others, critical thinking also covers thinking about the consequences that result from new technology. This sounds quite familiar compared to the issue of digital accessibility, where it is also critical to think about the implications technologies have for marginalized groups such as people with a disability. Over the past two decades, further literacy concepts have been developed, such as "media literacy" [10], "data literacy" [12], "information literacy" [1] or - most recently - "AI literacy" [9].

Central to literacy concepts is the individual's ability to encode and decode. In a narrower sense, a person should be able to read (decoding) and write (encoding) written language. These two basic skills and processes can also be applied to other contexts of literacy concepts, as exemplified by the example of "data literacy" [12]:

- Coding
    - Establishing a data culture
    - Providing data
    - Evaluating data
- Decoding
    - Interpreting results
    - Interpreting data
    - Deriving action

As a rule, the competencies operationalized in this way are specified at different levels of complexity and thus outline basic, advanced, and expert levels.

## 4 Draft of Digital Accessibility Literacy

As with the other literacies, digital accessibility literacy could be defined as the set of skills and knowledge needed to design digital content and technologies so that they can be found, accessed, and used by all people in the usual way, without particular difficulty, and usually without outside help. This includes understanding and applying principles and techniques for creating accessible digital content and services (encoding), as well as being able to interpret and use the accessibility of content (decoding). "Digital Accessibility

Literacy includes the integration of awareness raising, technical standards, inclusive design practices, and the consideration of user feedback and adaptation to improve the usability and accessibility of digital resources continuously.

An initial concept for such "digital accessibility literacy" along the two strands of encoding (creating digital accessibility) and decoding (understanding, evaluating, and using digital accessibility) could look as follows:

- Encoding (creating digital accessibility)
    - Understanding and promoting digital accessibility: Raising awareness, fostering a culture of digital accessibility and inclusion.
    - Creating accessible content: Understanding and applying design principles, adhering to technical standards.
    - Use and develop accessible technologies: use software and technologies and develop innovative solutions.
- Decoding (understanding and using digital accessibility)
    - Interpreting accessible digital content: using and understanding accessible presentation, enabling access to information
    - Obtain and use feedback: Obtaining feedback from users, using feedback for adaptations, and making adaptations
    - Develop inclusive solutions: Using knowledge, feedback, and data to improve digital accessibility, developing and implementing strategies

The concept of "digital accessibility literacy" can be used as a starting point for developing an educational culture of digital accessibility. As a basis for a discourse on the necessary skills, their acquisition, and teaching, the concept can help to address not only technical understanding but also ethical and social responsibility, which can create an educational culture in which digital accessibility is embedded as a natural part of all digital interactions and developments.

## 5 Discussion and Conclusion

What are the benefits of using the concept of Digital Accessibility Literacy (DAL)? In addition to more general benefits such as inclusivity awareness, skill development, or empathy building, distinct benefits set this concept apart.

First, a concept such as Digital Accessibility Literacy can help address the problem of "best practice" as described by [7]. Current practice in teaching accessibility is often characterized by "high-agreement, high-certainty territory of standards, guidance and monitoring of best practice." Everybody does the same because they see their colleagues doing it this way. A well-known example is teaching lecturers how to make their Word documents and PowerPoint presentations accessible. But do lecturers have to learn this, or should they instead learn about understanding and using digital accessibility, the "Decoding" aspect of Digital Accessibility Literacy? As a result of such a change, lecturers might change their way of creating teaching and learning materials, resulting in more accessible and flexible materials.

Second, digital accessibility literacy can promote interdisciplinary applications and collaboration between academic disciplines. Computer science is still the dominant field in digital accessibility. And much of the teaching of accessibility is about adhering to technical standards and norms. Pedagogy and other disciplines may have something to add but often fall short when implementing digital accessibility. The concept of Digital Accessibility Literacy can be used as a blueprint for discussing current approaches to teaching accessibility in various disciplines and developing innovative and unified ways of learning about accessibility.

How does the concept of DAL relate to the Accessibility Knowledge Area (ARA) proposed in [2]? Overall, the concept of DAL is not yet as far developed as ARA. At the moment, DAL is a draft and primarily useful as an input for discussion and as a basis for further developments, while the ARA is very differentiated. However, both approaches share some commonalities when it comes to learning about digital accessibility. Both approaches recognize the need for accessibility awareness, the ability to create accessible digital assets, and knowledge of assistive technologies. Both also acknowledge the need to be able to understand and use accessible digital assets, to get feedback from users with disabilities, and to be able to keep learning about accessibility. While the ARA strives to enhance the computer science curriculum, the DAL aims to be more broadly useful. Today, the responsibility for digital accessibility does not lie solely with computer scientists and IT specialists. Any professional who uses IT to create and use digital assets also needs to know about digital accessibility. Therefore, it may be helpful to find an approach to include digital accessibility in the curriculum for future teachers, lecturers, designers, or even data scientists. Both approaches do not address the question in which order the various topics should be learned. What is best in order to become an expert in the field of digital accessibility? Should I become aware of the need for accessibility first? Or should I become knowledgeable about evaluating the accessibility of existing digital assets before creating accessible ones myself? One strength of DAL is the accentuation of two complementary perspectives – to become an expert in the field of digital accessibility, I need to be able to both encode and decode digital accessibility, and I need to do this in a certain context. In contrast, the five components of ARA appear to stand side by side but it is unclear whether and where they are connected with each other.

The development of the concept of Digital Accessibility Literacy is still in an early stage. It needs further refinement and practice examples of implementing the various aspects and facets. However, this is the third benefit of this concept, as it can help build a pedagogical culture of teaching (and learning) digital accessibility in various disciplines. It can be used to develop a shared understanding of what it takes to learn about digital accessibility and what is needed to teach it. As a thinking tool and a shared vocabulary, it can foster a dialogue between disciplines and between teaching and research. This can help promote extended professionalism in the field.